\documentclass{article}
\PassOptionsToPackage{numbers,sort&compress}{natbib}
\usepackage[final]{neurips_2026}

\makeatletter
\renewcommand{\@noticestring}{NeurIPS 2026 Workshop on On-Device Intelligence: Foundation Models under Real-World Constraints.}
\makeatother

\usepackage[utf8]{inputenc} 
\usepackage[T1]{fontenc}    
\usepackage{hyperref}       
\usepackage{url}            
\usepackage{booktabs}       
\usepackage{amsfonts}       
\usepackage{nicefrac}       
\usepackage{microtype}      
\usepackage{xcolor}         
\usepackage{kotex}
\usepackage{amsmath}
\usepackage{amssymb}
\usepackage{graphicx}
\usepackage{subcaption}
\usepackage{multirow}

\title{Tap2Listen: Instant Target Switching for Streaming Audio-Visual Speaker Extraction}  

\author{%
  Hansol Park, Minkyung Song, Zinu Kim, Yujun Lee, \\
  \textbf{Hoseong Ahn, Hyunjin Choi, Kyuhong Shim} \vspace{0.05cm} \\
  Sungkyunkwan University \vspace{0.05cm} \\
  { \texttt{firstri@skku.edu, khshim@skku.edu}}
}

\begin{document}

\maketitle

\begin{abstract}
Streaming audio-visual target speaker extraction (AV-TSE) uses a target speaker's face to extract the speaker's speech from an input audio mixture in real time.
However, existing systems fix the target speaker for an entire clip, while conversation demands frequent and immediate target switches.
To address this problem, we propose Tap2Listen, a streaming AV-TSE model designed for instant response to target switches.
The model separates computation by modality: the target-agnostic audio encoder processes only the mixture, while the target-specific visual cues of the selected face condition only the decoder.
A switch requires only a $0.12$\,ms re-projection and takes effect immediately.
We also design component-wise losses: a prediction loss for the encoder and a supervised extraction loss for the decoder.
For evaluation, we measure how quickly output quality recovers after a switch.
On the standard LRS3-2mix test set, Tap2Listen improves SI-SNRi from $8.85$ to $9.68$ dB at a comparable parameter count. 
After a switch to a known face, the model produces full-quality output from the first frame, whereas the baseline needs about 1 second to recover.
Finally, we show that the proposed streaming AV-TSE model runs in real time on a single CPU core.
\end{abstract}

\section{Introduction}\label{sec:intro}

\begin{figure}[t!]
     \centering
     \includegraphics[width=0.9\linewidth]{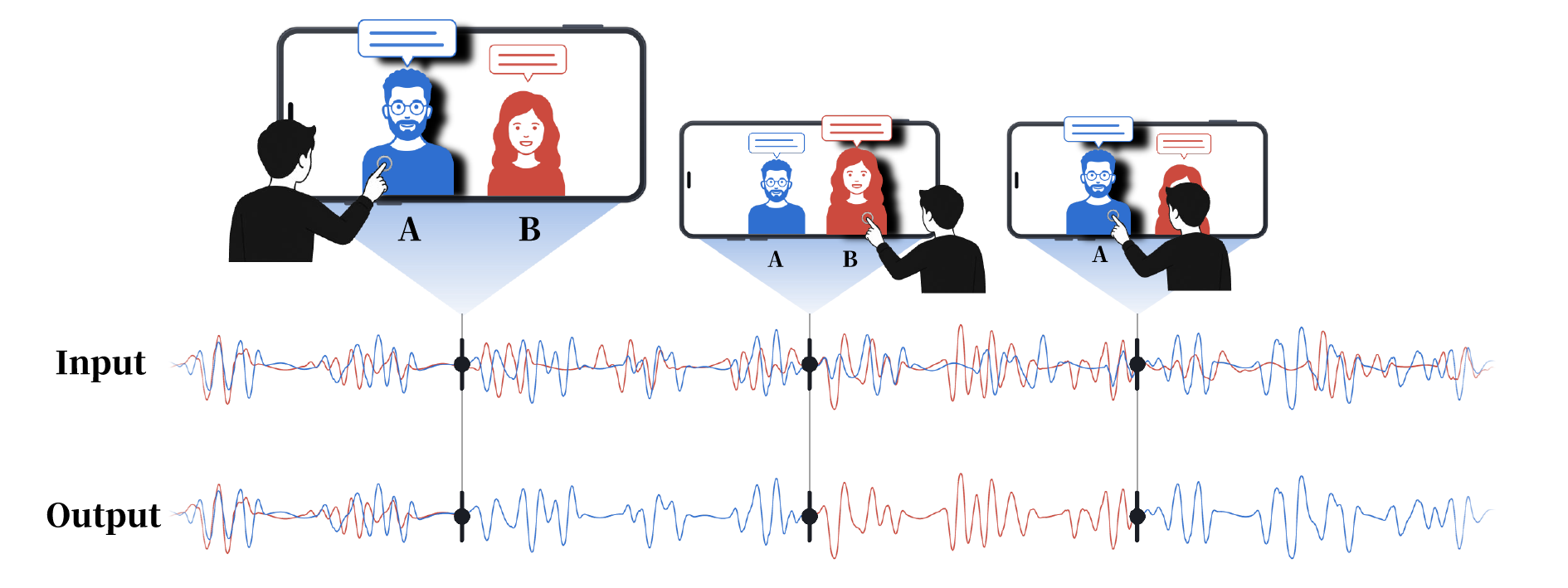}
     \caption{Tap2Listen interaction. Two speakers are present in the same mixture; tapping face A or B immediately switches the output to the selected speaker.}
     \label{fig:intro}
 \end{figure}

As hearables and smart glasses gain cameras and on-device computation, users can directly choose whom to listen to~\cite{veluri2023semantic, icassp/VeluriCICYG23, veluri2024lookonce}.
Audio-visual target speaker extraction (AV-TSE) uses face information to extract a single speaker's speech from a mixture~\cite{icassp/PanTX021,ochiai19_interspeech}.
Recent causal streaming AV-TSE systems achieve strong extraction quality under real-time playback constraints~\cite{pan25avskim, yu2025avcrossnet,  chen2024rtlavoce}.
However, these systems are typically designed for a fixed target speaker~\cite{chen2024rtlavoce,yu2025avcrossnet,sang2025swiftnet}.
In practice, users in a conversation switch targets repeatedly and expect each switch to take effect immediately.
For example, tapping a face on the screen should leave only the speaker audible, and tapping a different face should switch the output to the new speaker with minimal delay (Fig.~\ref{fig:intro}).
Nevertheless, interactive target switching has received little attention.

A key difficulty comes from two aspects of existing AV-TSE systems.
First, audio and visual information are often tightly coupled throughout processing, making a change of visual target entangled with ongoing audio computation.
Second, streaming systems may accumulate target-dependent state over time.
After a target change, information associated with the previous speaker can persist in recurrent, autoregressive, or speaker-tracking states, delaying the transition to the new target~\cite{li2025memo,li2025momuse,pan25avskim}.
Together, these properties make instant target switching difficult and motivate an architecture that separates causal and persistent audio processing from target-specific conditioning.

\begin{figure*}[t]
     \centering\includegraphics[width=1.0\linewidth]{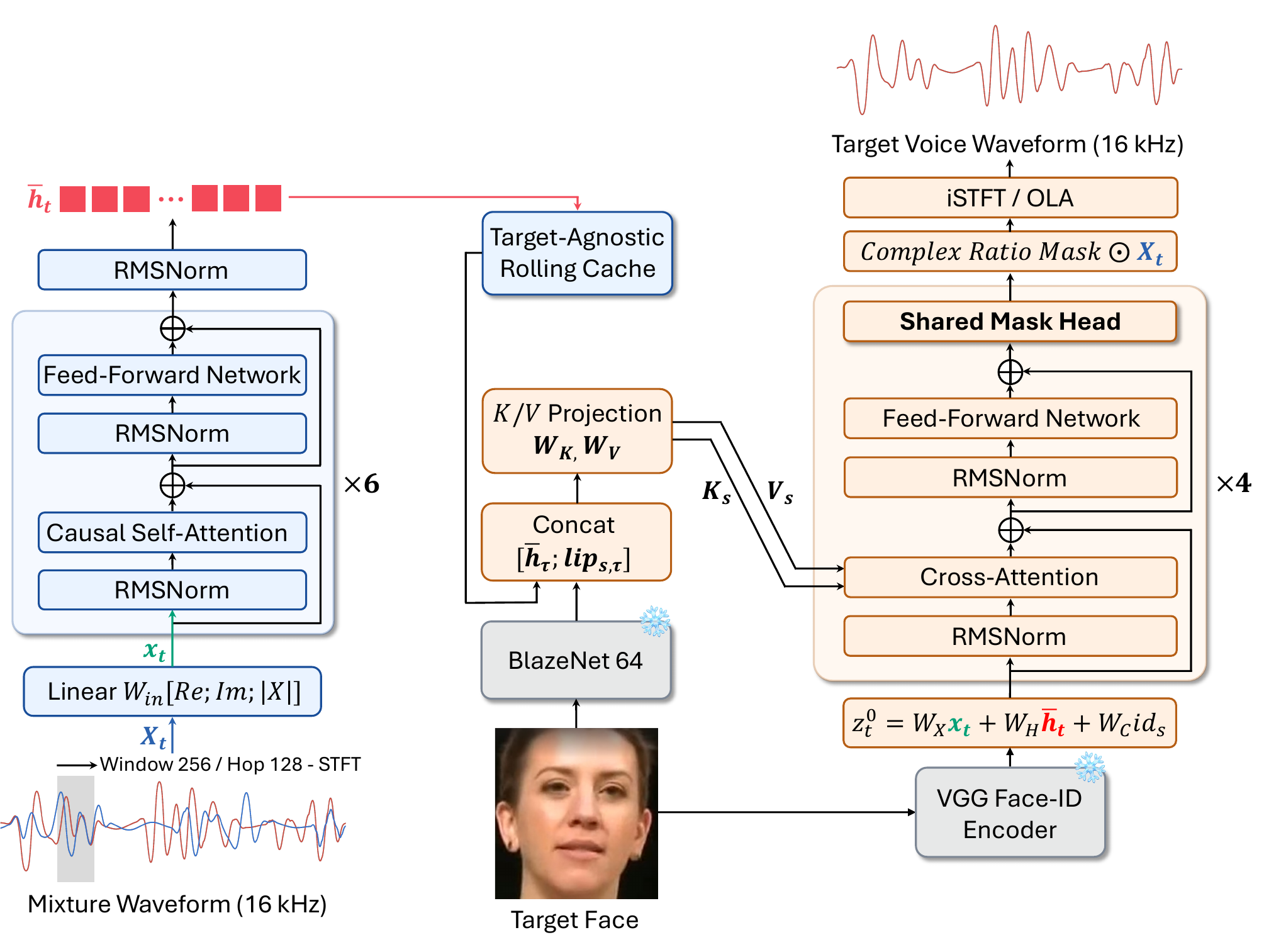}
     \caption{
     Tap2Listen architecture.
     Blue denotes target-agnostic computation that is retained across target switches.
     Orange indicates target-specific computation recomputed at each switch.
     Gray blocks are frozen pretrained networks.
     }
     \label{fig:method}
\end{figure*}

We propose Tap2Listen, a streaming AV-TSE model designed for instant target switching.
The key idea is to separate target-agnostic and target-specific computation.
A causal audio encoder processes only the audio mixture and maintains a shared temporal cache, while visual cues from the selected face condition only the decoder (Fig.~\ref{fig:method}).
Switching targets does not invalidate the encoder cache and requires only re-projecting the cached memory with the newly selected face.
As a result, a ``tap'' recomputes only the cached-window projection, immediately producing the new speaker's output.

We evaluate both conventional extraction quality and the switching transition itself, using response delay and post-switch recovery curves.
On LRS3-2mix~\cite{LRS3}, Tap2Listen achieves $9.68$\,dB SI-SNRi compared with 8.85 dB of the baseline~\cite{pan25avskim} at a comparable parameter count.
More importantly, the model reaches full quality from the first frame after switching to a cached face, while the baseline requires about one second to recover.

\noindent Our contributions are summarized as follows.
\begin{itemize}
    \item We formulate interactive target switching as a distinct requirement of streaming AV-TSE and evaluate the transition using response delay and post-switch recovery, rather than relying only on utterance-level averages.
    \item We propose Tap2Listen, which keeps persistent audio processing target-agnostic and decouples target-specific computation from the streaming audio state, together with component-wise extraction and prediction losses for training the decomposed architecture.
    \item We achieve competitive extraction quality, immediate switching to cached targets without a recovery period, and real-time execution on a single mobile CPU core with Tap2Listen.
\end{itemize}

\section{Tap2Listen: Instant Target Switching for Streaming AV-TSE}\label{sec:method}

\subsection{Architecture}

Tap2Listen consists of a target-agnostic causal audio encoder that processes only the mixture and a target-conditioned decoder that uses the selected face, with $8.06$\,M trainable parameters (Fig.~\ref{fig:method}).
The encoder projects $16$\,ms-window, $8$\,ms-hop STFT frames to $256$ dimensions, followed by $6$ pre-RMSNorm~\cite{NEURIPSrms} blocks with local causal self-attention~\cite{transformer} over $128$ frames.
Because the encoder is target-agnostic, the encoder cache remains valid across target switches.
Frozen visual frontends provide lip features from causal BlazeNet64 at $25$\,Hz~\cite{pan25avskim} and face ID vectors from a face embedder~\cite{VGGFace2}.
A rolling buffer stores each face's lip features over the same $4$\,s window.

The decoder is the only target-dependent component.
For a selected face $s$, temporally aligned lip features are combined with the cached audio representation to form the attention memory,
\begin{equation}
k_\tau,\, v_\tau = W_{K,V}\big[\,\bar{h}_\tau \,;\, \mathrm{lip}_{s,\tau}\,\big],
\qquad \tau \in [\,t-M,\, t\,]
\label{eq:mem}
\end{equation}
where $M=500$ frames ($4$\,s at the $8$\,ms hop), making a tap $1.2\times10^{8}$ MACs.
The face identity embedding contributes the decoder query:
\begin{equation}
z_t^{(0)} = W_X x_t + W_H \bar{h}_t + W_C \mathrm{id}_s
\label{eq:query}
\end{equation}
The time-varying, audio-synchronized lip features~\cite{Chung_2017_CVPR} provide visual cues for cross-attention to guide the temporal correspondence between lip motion and acoustics, whereas the fixed identity embedding appears only in the decoder query.

The query passes through $K=4$ cross-attention and FFN blocks with LayerScale~\cite{iccv/TouvronCSSJ21}.
A shared mask head produces a complex ratio mask at each block, from which the iSTFT with overlap-add reconstructs the waveform at every $8$\,ms hop.
The decoder uses neither self-attention over previous decoder outputs nor autoregressive feedback; causality is preserved because its memory contains only frames up to the current time.

\subsection{Training Strategy and Loss}
\label{sec:loss}

During training, each speaker in the mixture is treated in turn as the target.
The encoder runs once for each mixture, while the memory projection is recomputed for each speaker.
The shared mask head outputs a mask at each decoder block $k$, from which we obtain the target estimate $\hat{s}^{(k)}$ under deep supervision~\cite{aistats/LeeXGZT15}.
The extraction loss is:%
\begin{equation}
    \mathcal{L}_{\text{ext}} = \sum_{s}\sum_{k} w_k \cdot
    \mathrm{clamp}\big(-\text{SI-SDR}(\hat{s}^{(k)}_s),\, -30\,\text{dB}\big)
    \label{eq:loss_ext}
\end{equation}
where $w=(0.1,0.2,0.3,0.4)$ assigns larger weights to later decoder blocks.
Each term is capped at $-30$\,dB~\cite{icassp/RouxWEH19}, so examples already above $30$\,dB SI-SDR contribute no gradient.
To encourage the target-agnostic encoder to capture predictive temporal structure, we attach a linear auxiliary head on the top-layer encoder state to predict the change in log magnitude, $\Delta\log|X|$, five frames ahead:
\begin{equation}
    \mathcal{L}_{\text{pred}} = \sum_{t} \Big\|\, g(\bar{h}_t)\,-\, \Big(\log |X|_{t+5} - \log |X|_{t}\Big) \,\Big\|_2^2
    \label{eq:loss_pred}
\end{equation}
This objective uses only the mixture signal and requires no target-speaker labels.
The final objective is a weighted sum of two losses: $\mathcal{L} = \mathcal{L}_{\text{ext}} + \lambda\mathcal{L}_{\text{pred}}$, with $\lambda=0.1$.

\subsection{Inference and Target Switching}
\label{sec:inference}
At inference, the encoder state is retained in a shared cache.
A switch from $s$ to $s'$ requires only recomputing Eq.~(\ref{eq:mem}) over the cached frames with the buffered lip features of $s'$, and replacing the identity term in Eq.~(\ref{eq:query}), without re-running the audio encoder.
Because there is no target-specific cache or state, returning to a previously selected face costs exactly the same target-specific recomputation as any other switch.
In contrast, streaming systems that rely on autoregressive feedback or speaker-embedding momentum inherit state from the previous target after a switch and must wait for it to decay~\cite{pan25avskim,li2025momuse,li2025memo}.
\begin{table*}[t!]
    \centering
    \caption{Standard LRS3-2mix test (3,000 fixed mixtures). SI-SNRi and SNRi are reported in dB.}
    \setlength{\tabcolsep}{5pt}
    \resizebox{1.0\linewidth}{!}{
    \begin{tabular}{l|ccc|cccc}
        \toprule[1.2pt]
        \multirow{2}{*}{Model} & \multicolumn{3}{c|}{Setup} & \multicolumn{4}{c}{Improvement} \\
        \cmidrule{2-4} \cmidrule{5-8}
         & Params~$\downarrow$ & Training budget~$\downarrow$ & Alg.\ lat.~$\downarrow$ & SI-SNRi~$\uparrow$ & SNRi~$\uparrow$ & PESQi~$\uparrow$ & STOIi~$\uparrow$ \\
        \midrule[1.2pt]
        AV-SkiM~\cite{pan25avskim}
          & 8.1\,M & 4$\times$V100, 150\,ep & $\sim$1\,ms
          & 8.85 & 9.16 & 0.52 & 0.19 \\
        \textbf{Tap2Listen (ours)}
          & 8.1\,M & 1$\times$3080, 150\,ep & 16\,ms
          & \textbf{9.68} & \textbf{10.05} & \textbf{0.71} & \textbf{0.20} \\
        \bottomrule[1.2pt]
    \end{tabular}}
    \label{tab:standard}
\end{table*}

\section{Experimental Results}\label{sec:exp}
\subsection{Setup}\label{sec:setup}

\paragraph{Training and Evaluation.}
We trained the model for $150$ epochs ($28$~GPU-h) on a single NVIDIA RTX 3080.
Training uses bf16 precision, a batch size of $16$, AdamW~\cite{iclr/LoshchilovH19} with a cosine schedule from $10^{-3}$ to $10^{-5}$.
We use an exponential moving average (EMA) of the model parameters with a decay of $0.999$ and report all results using the EMA weights.
For comparison, we use the released AV-SkiM~\cite{pan25avskim} checkpoint and re-evaluate it in our environment using the same data and evaluation pipeline as Tap2Listen.
We report SI-SNRi, SNRi, PESQi, and STOIi for extraction quality, and measure selection-to-audio delay and post-switch SI-SDR recovery for target switching.
\paragraph{Dataset.}\label{sec:data} %
We use the ClearerVoice LRS3-2mix protocol with $40$k/$5$k/$3$k train/val/test rows from $66{,}974$ LRS3~\cite{LRS3} utterances, with only the test split being speaker-independent.
Nothing is pre-mixed. The target is anchored at $0$\,dB, and the power-matched interferer is mixed at $\pm10$\,dB SNR. In training, the interferer and SNR are redrawn from the $58.5$k-utterance pool, so $40$k denotes anchors, not mixtures.
Switching uses our reconstruction of the AV-SkiM attention-shift protocol, splitting the same $3{,}000$ test rows $1{,}014$/$1{,}012$/$974$ no-switch/switch-face/new-face.
The released checkpoint reproduces no-switch and switch-face means within $0.10$ and $0.56$\,dB, respectively; since the original new-face seed is unavailable, all models are compared on the same reconstructed mixtures.

\subsection{Fixed-Target Extraction}\label{sec:exp-qualit}

Tap2Listen achieves $9.68$\,dB SI-SNRi, 0.71 PESQi~\cite{pesq-quality}, and 0.20 STOIi~\cite{STOI}, exceeding the baseline's numbers under identical conditions (Table~\ref{tab:standard}).
This corresponds to a $+0.83$\,dB improvement at matched parameter count while being trained on a single RTX 3080 rather than four V100 GPUs.
These results show that Tap2Listen retains strong fixed-target extraction performance despite being designed primarily for interactive target switching. 
Tap2Listen has a 16 ms algorithmic latency, compared with approximately 1 ms for AV-SkiM.
This difference mainly reflects the processing granularity: Tap2Listen operates in the STFT domain, whereas AV-SkiM operates directly on the waveform.
While the latter provides finer algorithmic granularity, it requires substantially more frequent model invocations; Tap2Listen instead processes one frame every 8 ms, reducing the frequency of model invocations for practical on-device execution.

\begin{figure*}[t!]
     \centering\includegraphics[width=0.95\linewidth]{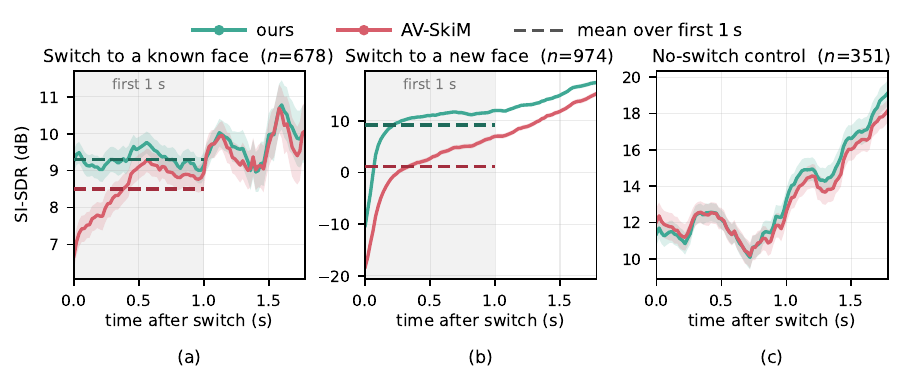}
     \vspace{-0.3cm}
     \caption{SI-SDR after switching, measured in $200$\,ms windows (mean $\pm$ s.e.m.). (a) known-face, (b) new face, and (c) no-switch control with matched pseudo-switch offsets. Shading marks the first $1$\,s after the switch, and dashed lines show the mean over this interval.}
     \label{fig:recovery}
\end{figure*}

\begin{table}[t]
  \centering
  \caption{SI-SNRi (dB) of the switching protocol. Higher is better. Avg. denotes the full-utterance average, while $\le1$\,s and $>1$\,s denote the mean over $200$\,ms windows after the switch.}
  \vspace{0.1cm}
  \setlength{\tabcolsep}{11pt}
  \resizebox{1.0\linewidth}{!}{
  \begin{tabular}{l|cccccccc}
    \toprule[1.2pt]
    & \multirow{2}{*}{No switch} & \multicolumn{3}{c}{Known face} & \multicolumn{3}{c}{New face} \\
    \cmidrule{3-8}
    & & Avg. & $\leq 1s$ & $> 1s$ & Avg. & $\leq 1s$ & $> 1s$ \\
    \midrule[1.2pt]
    AV-SkiM~\cite{pan25avskim}
      & \textbf{9.56}
      & 8.05 & 8.50 & 9.63
      & 4.73 & 1.08 & 10.82 \\
    \textbf{Tap2Listen (Ours)}
      & 9.51
      & \textbf{9.08} & \textbf{9.31} & \textbf{9.79}
      & \textbf{8.16} & \textbf{9.16} & \textbf{14.38} \\
    \bottomrule[1.2pt]
  \end{tabular}}
  \label{tab:switch}
\end{table}

\subsection{Target Switching}\label{sec:exp-switch}

Without a target change, the two systems perform comparably: AV-SkiM reaches $9.56\,$dB and Tap2Listen $9.51\,$dB, a $0.05\,$dB gap well within the between-utterance variation in Fig.~\ref{fig:recovery}(c). 
After switching to a known face, Tap2Listen leads by $2.59$\,dB already in the first $200$\,ms window with no recovery period, whereas AV-SkiM needs about $1$\,s to close the gap (Fig.~\ref{fig:recovery}).
For the first-appearing face, the gap reaches $\mathbf{8.08}$\,dB over the first $1$\,s after the tap (Table~\ref{tab:switch}), and peaks at $11.9$\,dB at $80$\,ms; the baseline still trails by $3.0$\,dB at $1.6$\,s.
In contrast, the corresponding full-utterance gains are only $1.03$\,dB for a known face and $3.43$\,dB for a first-appearing face, indicating that utterance-level averages substantially obscure the transition gap.

\subsection{On-device Latency}\label{sec:exp-device}

We export the streaming path as a static-shape ONNX graph and evaluate it on a Galaxy S22 (Snapdragon 8 Gen 1) using ONNX Runtime CPU with fp32, single-thread execution on a fixed big core.
After a five-minute thermally settled run, Tap2Listen processes each 8\,ms frame in $3.2$\,ms ($40$\% of the $8$\,ms budget) at p95, with no over-budget frame.
A tap-triggered re-projection takes $5.2$\,ms at p95, giving $8.5$\,ms transition latency including the current frame computation.
Because regular frames retain substantial timing margin, the additional tap computation is absorbed without accumulating delay.
Larger blocks reduce computation but increase buffering delay, making $8$\,ms the minimum-delay operating point.
The reported latency covers the full streaming path, including STFT/iSTFT, frame buffering, encoder and cache updates, target-conditioned mask estimation, and overlap-add reconstruction.
\section{Conclusion}\label{sec:conclusion}

We proposed Tap2Listen, a streaming AV-TSE model for interactive target switching that separates persistent target-agnostic audio processing from target-conditioned decoding.
Tap2Listen confines all target-dependent computation to a single memory projection, so each switch is one re-projection burst ($5.2$\,ms at p95 on a Galaxy~S22) whose output matches extraction conditioned on the new speaker from the start to $\approx-70$\,dB, with no state-driven recovery period.
Beyond conventional extraction metrics, we evaluated switching through response delay and post-switch recovery; Tap2Listen achieves $9.68$\,dB SI-SNRi on LRS3-2mix and substantially improves transition behavior over AV-SkiM.
The streaming path also operates within the real-time budget of a single mobile CPU core.
These results establish target switching as a distinct systems property of streaming AV-TSE that should be explicitly designed and evaluated at the transition itself.

\begin{ack}
This work was supported by the National Research Foundation of Korea (NRF) grant funded by the Korea government (MSIT) (RS-2026-25602529). This work was also supported by the Institute of Information \& Communications Technology Planning \& Evaluation (IITP) grants funded by the Korea government (MSIT) (IITP-2026-RS-2026-25615334, AI Star Fellowship Support Program; RS-2019-II190421, Artificial Intelligence Graduate School Program (Sungkyunkwan University)).
\end{ack}

{
    \small
    \bibliographystyle{plain}
    \bibliography{reference}
}


\newpage
\appendix
\section{Related Work}\label{app:related}
\subsection{Non-Streaming Methods}
\label{ssec:related_nonstream}

AV-TSE methods use visual cues from on-screen speakers to estimate the target speech from a mixture.
Looking to Listen~\cite{ephrat2018looking} takes the visual stream of each detected face and the mixed audio as input and separates the speech corresponding to each face.
The model uses visual information to improve separation performance and to associate each separated speech signal with the corresponding visible speaker.
The authors note that the method does not operate in real time and is better suited to video post-processing.

Subsequent studies distinguish the roles of visual cues more explicitly or reuse target speech obtained during separation as an additional speaker cue.
VisualVoice~\cite{gao2021visualvoice} uses two distinct visual cues: lip motion and facial appearance.
Lip motion provides visual information temporally aligned with the current utterance, while facial appearance from a single image serves as a cue to the speaker's vocal characteristics.
VisualVoice jointly learns speech separation and cross-modal face-voice representations to model this relationship.
MuSE~\cite{icassp/PanTX021} reuses target speech from the preceding extraction stage.
It generates a self-enrolled speaker embedding from this speech and visual features and uses the embedding to condition mask estimation in the next extractor block.
Repeating this process across extractor blocks progressively refines the target speech.

Alongside these developments, other studies improve architectures for fusing and processing audio and visual information.
CTCNet~\cite{li2024ctcnet} extracts audio and visual features at multiple temporal resolutions and repeatedly fuses them at multiple processing stages.
The model feeds the fused features back into subsequent audio and visual feature extraction, so each modality influences later stages.
RTFS-Net~\cite{pegg2024rtfsnet} represents audio in the time-frequency domain, models the temporal and frequency dimensions separately, and fuses visual features with the audio representation through attention.
It also reduces computational cost by performing its main computations on representations with fewer channels and lower time-frequency resolution.
These methods evaluate separation performance with a fixed target speaker in non-streaming settings and do not evaluate user-initiated target switching during processing.

\subsection{Streaming Methods}
\label{ssec:related_stream}

Streaming methods remove or limit dependence on future inputs to operate in real time.
RT-LA-VocE~\cite{chen2024rtlavoce} uses causal audio and visual encoders, an Emformer, and a causal vocoder to reduce algorithmic latency to one video frame ($40$\,ms) without future inputs.
Online AV-CrossNet~\cite{yu2025avcrossnet} uses one-frame look-ahead for online processing, while RAVEN~\cite{ma2025raven} uses a pretrained visual encoder with two-frame look-ahead and runs in real time on a CPU.
Swift-Net~\cite{sang2025swiftnet} uses causal convolutions and a unidirectional recurrent module in a lightweight architecture that operates without future inputs.

Some streaming methods reuse information from earlier segments to guide ongoing target speaker extraction.
MoMuSE~\cite{li2025momuse} stores past voiceprints of the target speaker in a memory bank and combines them with current speaker information to track the same speaker when visual cues are unreliable.
However, the authors evaluate MoMuSE under corrupted or missing visual cues for a fixed target speaker and do not evaluate target switching during streaming.
The algorithmic or processing latencies reported in prior streaming studies refer to delays during continuous processing with a fixed target speaker, not the delay until the output responds to a new target speaker after a target switch.

Pan et al.~\cite{pan25avskim} directly evaluate target speaker switching.
They switch the target at an arbitrary point while two speakers speak simultaneously and measure SI-SNRi before and after the switch.
The authors also argue that aggregating past separated signals into a single speaker embedding can create momentum that hinders rapid target switching.
However, average SI-SNRi before and after the switch does not show how separation performance changes immediately after the switch.
The proactive hearing assistant~\cite{hu2025proactive} measures SI-SDRi over successive intervals after a conversation partner starts speaking, but does not use explicit target selection as the reference point.
We use the target-selection time as the reference point and measure both the delay until the output responds to the new target speaker and the temporal evolution of separation performance immediately after the switch.




\end{document}